\documentclass[a4paper,11pt]{article}
\usepackage{pos}
\usepackage{booktabs,tabulary,slashed}

\newcommand{\A}{\mathcal{A}}
\newcommand{\BR}{\mathcal{B}}
\renewcommand{\O}{\mathcal{O}}
\newcommand{\M}{\mathcal{M}}
\newcommand{\F}{\mathcal{F}}
\newcommand{\G}{\mathcal{G}}
\renewcommand{\H}{\mathcal{H}}
\newcommand{\keV}{\,{\rm keV}}
\newcommand{\MeV}{\,{\rm MeV}}
\newcommand{\GeV}{\,{\rm GeV}}
\renewcommand{\Im}{{\rm Im}\,}
\renewcommand{\Re}{{\rm Re}\,}
\newcommand{\ep}{\eta^{(\prime)}}
\newcommand{\diff}{\text{d}}
\newcommand{\munu}{{\mu\nu}}
\newcommand{\etap}{\eta^{(\prime)}}

\def\vev#1{\big\langle #1 \big\rangle}

\def\vecsign{\mathchar"017E}
\def\dvecsign{\smash{\stackon[-3.55pt]{\vecsign}{\rotatebox{180}{$\vecsign$}}}}
\def\dvec#1{\def\useanchorwidth{T}\stackon[-4.2pt]{#1}{\,\dvecsign}}
\usepackage{stackengine}
\stackMath

\title{$C$ and $CP$ violation in effective field theories and applications to $\eta$-meson decays}
\ShortTitle{$C$ and $CP$ violation in effective field theories and applications to $\eta$-meson decays}

\author*[a]{Bastian Kubis}

\affiliation[a]{Helmholtz-Institut für Strahlen- und Kernphysik (Theorie) and
Bethe Center for Theoretical Physics, Universität Bonn, 53115 Bonn, Germany}

\emailAdd{kubis@hiskp.uni-bonn.de}

\abstract{The quest for sources of the simultaneous violation of $C$ and $CP$ symmetry was popular in the 1960s, but has since been neglected for a long time. We revisit the operators that break $C$ and $CP$ for flavor-conserving transitions in both the Standard Model effective field theory and the low-energy effective field theory, which subsequently can be matched to light-meson physics using chiral perturbation theory. As applications, we discuss in particular the $C$-odd Dalitz plot asymmetries in $\eta\to3\pi$, but also decays with dilepton pairs in the final state, such as long-distance contributions to the rare semileptonic decays $\eta\to\pi^0\ell^+\ell^-$ as well as asymmetries in $\eta^{(\prime)} \to \gamma \ell^+\ell^-$ and $\eta^{(\prime)} \to \pi^+\pi^-\ell^+\ell^-$.}

\FullConference{The 11th International Workshop on Chiral Dynamics (CD2024)\\
26--30 August 2024\\
Ruhr University Bochum, Germany\\}

\begin{document}
\maketitle

\section{Patterns of discrete symmetry breaking and \boldmath{$\eta$ and $\eta'$} mesons}
The lightest flavor-neutral mesons, 
the $\eta$ and the $\eta'$, are ideal labs for a plethora of symmetry tests.
With the exception of parity, they have the quantum numbers of the vacuum, $I^G(J^{PC}) = 0^+(0^{-+})$.  
The lighter of the two, the $\eta$ with $M_\eta = 547.86\MeV$,
is largely a (pseudo-) Goldstone boson of spontaneously broken chiral symmetry.
The smallness of its width, $\Gamma_\eta=1.31\keV$, can be understood as many of its decay modes, strong or electromagnetic, are forbidden at leading order due to $P$, $C$, $CP$, isospin/$G$-parity, or angular momentum conservation~\cite{Nefkens:2002sa}.
The  heavier $\eta'$, $M_{\eta'} = 957.78\MeV$, is no Goldstone boson due to the $U(1)_A$ anomaly, however its width, $\Gamma_{\eta'}=196\keV$, is still much smaller than those of, e.g., the $\omega(782)$ or $\phi(1020)$ vector resonances of comparable masses.  

Decays of $\eta$ and $\eta'$ mesons can test physics both within the Standard Model (SM) and beyond.
For an overview of stringent SM tests, as well as searches for physics beyond the SM (BSM) in the form of
weakly-interacting new light particles (dark photons, protophobic or leptophobic gauge bosons of new $U(1)$ symmetries, light Higgs-like scalars, axion-like particles), we refer to the review Ref.~\cite{Gan:2020aco}  and references therein.
In these proceedings, we concentrate on tests of discrete symmetries, in particular searches for possible new sources of $CP$ violation.  As the $\eta$ and $\eta'$ mesons are $C$ and $P$ eigenstates, their decays are a flavor-conserving laboratory for such symmetry tests, with little or no SM background.
\begin{table}[b!]
\centering
\scalebox{0.9}{
\renewcommand{\arraystretch}{1.3}
\begin{tabular}{llll} 
\toprule
class     &  violated & conserved    & interaction \\
\midrule
0 & & $C$, $P$, $T$, $CP$, $CT$, $PT$, $CPT$ & strong, electromagnetic\\
I & $C$, $P$, $CT$, $PT$ & $T$, $CP$, $CPT$ & (weak, with no KM phase or flavor mixing) \\
II & $P$, $T$, $CP$, $CT$ & $C$, $PT$, $CPT$ & \\
III & $C$, $T$, $PT$, $CP$ & $P$, $CT$, $CPT$ & \\
IV & $C$, $P$, $T$, $CP$, $CT$, $PT$ & $CPT$ & weak \\
\bottomrule
\end{tabular}
\renewcommand{\arraystretch}{1.0}
}
\caption{Possible classes~I--IV of interactions that violate discrete spacetime symmetries, assuming $CPT$ invariance.
Table taken from Ref.~\cite{Gan:2020aco}.
 \label{tab:symbreaking}}
\end{table}
The different possible classes of violation and conservation of $C$, $P$, and $T$ (always assuming $CPT$ to be conserved) are enlisted in Table~\ref{tab:symbreaking}.  While the Standard Model weak interactions are in class~IV, violating all three discrete symmetries separately, they are in many circumstances close to class~I, violating $C$ and $P$ maximally, with $CP$ almost conserved.
The interactions of class~II, $P$-odd and $T$-odd, but conserving $C$, comprise the QCD $\theta$-term and other
possible higher-dimensional terms that are closely related to the physics of electric dipole moments (EDMs),
and therefore typically constrained so tightly that they will remain outside the reach of light-meson-decay
experiments in the foreseeable future.

In the rest of this article, we concentrate on class~III, $T$- and $C$-odd, but $P$-even interactions, which have been much less explored until rather recently.
As the flavor-neutral light pseudoscalars $\ep$, $\pi^0$ are eigenstates of $C$ with eigenvalues $C=+1$, any decay that involves these mesons only, together with an \textit{odd} number of photons, is directly a test of $C$-conservation.  Examples of such $C$-forbidden $\eta$ decays are listed in Table~\ref{tab:CVdecays}.
\begin{table}
  \centering
  \scalebox{0.9}{
\renewcommand{\arraystretch}{1.3}
\begin{tabular}{llll} 
\toprule
Channel     & Branching ratio   & Note & Ref. 
\\ 
\midrule
$\eta \to 3\gamma$     
&  $<1.6\times 10^{-5}$     
&  
& \cite{KLOE:2004ukf} 
\\ 
$\eta \to \pi^0\gamma$  
&  $<9\times 10^{-5}$ 
& Violates angular momentum conservation or gauge invariance
& \cite{Nefkens:2005ka} 
\\
$\eta \to \pi^0e^+e^-$       
&  $<7.5\times 10^{-6}$ 
&  $C$, $CP$-violating as single-$\gamma$ process 
& \cite{WASA-at-COSY:2018jdv} 
\\ 
$\eta \to \pi^0 \mu^+ \mu^-$ 
& $< 5\times 10^{-6}$ 
&  $C$, $CP$-violating as single-$\gamma$ process
& \cite{Dzhelyadin:1980ti} 
\\ 
$\eta \to 2\pi^0\gamma$ 
&  $<5\times 10^{-4}$
&
& \cite{CrystalBall:2005zrs} 
\\ 
$\eta \to 3\pi^0\gamma$ 
& $<6\times 10^{-5}$ 
&
& \cite{CrystalBall:2005zrs} 
\\  
\bottomrule
\end{tabular}
\renewcommand{\arraystretch}{1.0}
}
\caption{$C$-violating $\eta$ decay modes with experimental upper limits. Table adapted from Ref.~\cite{Gan:2020aco}.
\label{tab:CVdecays} }
\end{table}
However, searching for such rigorously $C$-forbidden decays may not be advantageous, as we expect all BSM
effects to be parametrically strongly suppressed, and any corresponding decay rates will hence be quadratically
small.  This is different for observables that are generated as interference effects between $C$-conserving
SM amplitudes and $C$-odd BSM physics, which will typically manifest themselves in the form of various
asymmetries.  We will discuss such asymmetry effects repeatedly in the following sections.

\section{\boldmath{$C$ and $CP$ violation for light mesons: a hierarchy of effective field theories}}\label{sec:C}

\begin{figure}
    \includegraphics*[width=\linewidth]{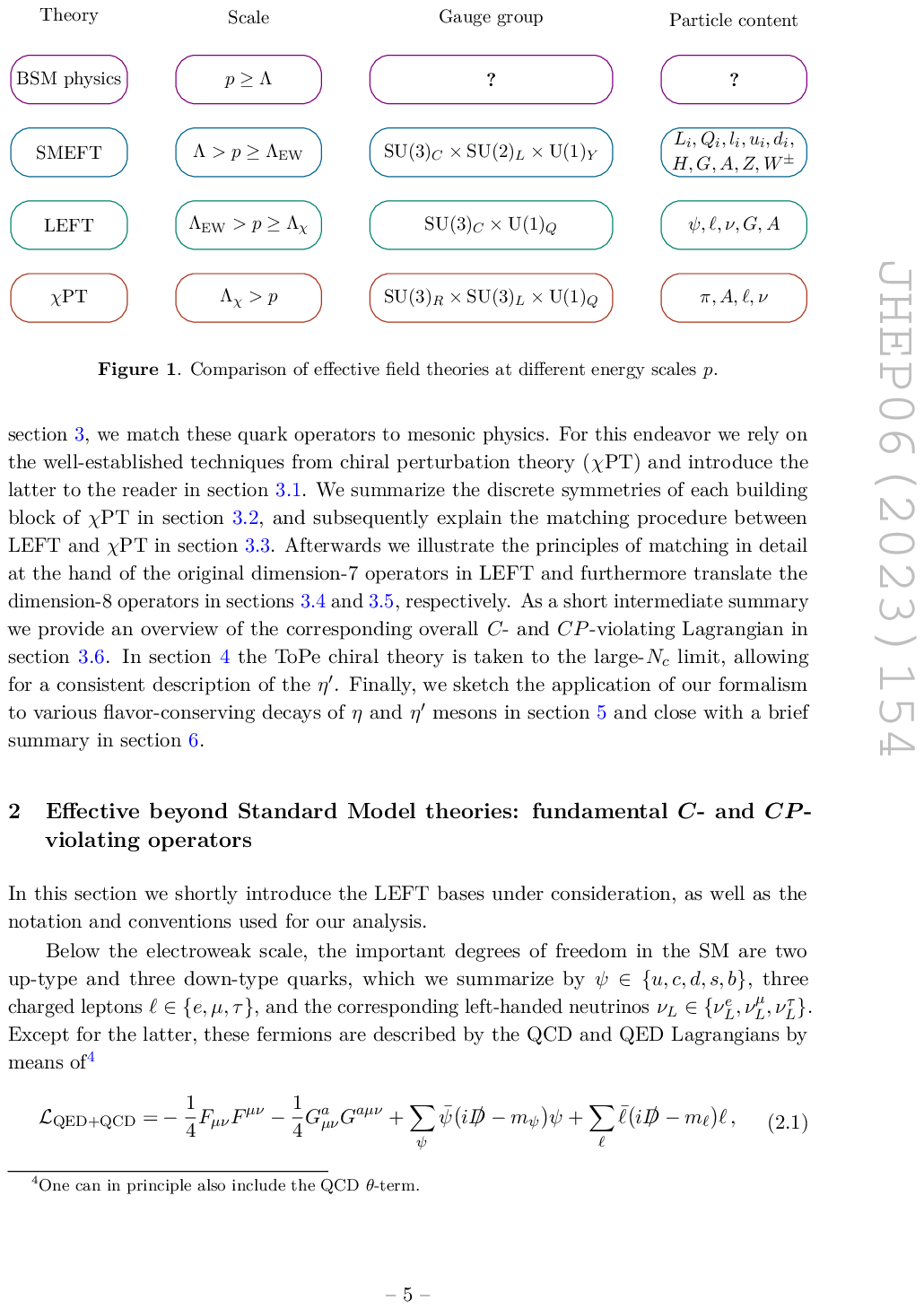}
    \caption{Hierarchy of different effective field theories operating at different scales,
      based on different gauge groups and involving different sets of degrees of freedom.
      Figure taken from Ref.~\cite{Akdag:2022sbn}. \label{fig:EFTs}}
\end{figure}

Effective operators on the quark level corresponding to class~III have already been written down in the
1990s. They appear at mass dimension seven and involve both four-quark or fermion--gauge-boson operators such as~\cite{Khriplovich:1990ef,Ramsey-Musolf:1999cub,Kurylov:2000ub}
\begin{equation}
\frac{1}{\Lambda^3} \bar{\psi}_f \gamma_5 \dvec{D}_\mu  \psi_f \, \bar{\psi}_{f^\prime} \gamma^\mu \gamma_5 \psi_{f^\prime} \, , \qquad \frac{1}{\Lambda^3} \bar{\psi}_f \sigma_{\mu\nu} \lambda_a \psi_f G^{\mu\lambda}_a F^\nu_\lambda \,. \label{eq:C-odd-ops}
\end{equation}
However, the theory framework in which these are to be understood has not been made very clear at the time.
As these operators involve fermion helicity flips as mediated by the Higgs vacuum expectation value (vev) $v$ in the SM, they are actually of mass dimension eight when formulated in terms of fields invariant under the SM symmetry group, the Standard Model effective field theory (SMEFT).
In other words, the scale $1/\Lambda^3$ in Eq.~\eqref{eq:C-odd-ops} ought to be re-interpreted rather as $v/\Lambda^4$ when matching the low-energy effective field theory (LEFT) below the electroweak scale to SMEFT.
In Ref.~\cite{Akdag:2022sbn}, where the hierarchy of effective theories for $C$-odd interactions has been
thoroughly investigated
from SMEFT via LEFT to the hadronic
realization in chiral perturbation theory, cf.\ Fig.~\ref{fig:EFTs},
it was therefore advocated to retain, in the LEFT framework, both chirality-breaking operators
of dimension seven scaling as $v/\Lambda^4$, and chirality-conserving operators of dimension eight
$\propto 1/\Lambda^4$.

An important loop-hole in this argument has been pointed out most recently in Ref.~\cite{Shi:2024yfa},
where it was demonstrated that the four-quark operator in Eq.~\eqref{eq:C-odd-ops} can also be generated
from a $W$-exchange diagram involving a dimension-6 $C$-odd $W\bar{q}q$ vertex.  While the resulting LEFT
operator is the same, it changes its scaling with the BSM scale and the Higgs vev according to
\begin{equation}
  \frac{v}{\Lambda^4} \longrightarrow \frac{v}{\Lambda^2} \frac{1}{M_W^2} \propto \frac{1}{v\Lambda^2} \,.
\end{equation}
In this way, $C$-odd interactions can be routed in dimension-6 SMEFT operators, and appear on a comparable
footing as the $P$-odd, $C$-even EDM operators beyond the $\theta$-term (such as quark EDMs etc.~\cite{deVries:2012ab,Bsaisou:2014oka}).  
This makes the search for these effects much more promising.
To what extent also the other LEFT operators investigated in Ref.~\cite{Akdag:2022sbn} can be thus promoted
to lower orders in the SMEFT counting remains to be investigated.

For the last step towards an effective theory usable directly in the decays of light mesons, the LEFT
operators, formulated in terms of quark, lepton, and massless gauge fields, needs to be matched to
chiral perturbation theory~\cite{Akdag:2022sbn}.  This is achieved in terms of the external-source method~\cite{Gasser:1983yg}:
LEFT operators are introduced as spurion fields, either violating chirality ($\propto \lambda^{(\dagger)}$)
in analogy to the way quark masses are introduced as explicit chiral symmetry breakers,
or conserving chirality ($\propto \lambda_{L,R}$), much as quark charges are introduced to take care
of the effects of hard photons~\cite{Urech:1994hd}.
An example for the hadronization of the leading (four-quark) LEFT operator into the chiral field $U$
is given by
\begin{equation}
  \frac{c_{\psi\chi}^{(a)}}{v \Lambda^2} \, \bar\psi \dvec D_\mu \gamma_5\psi\,\bar\chi\gamma^\mu\gamma_5\chi
    \longrightarrow 
\frac{c^{(a)}_{\psi\chi}}{{v\Lambda^2}}
i {g^{(a)}_1}\vev{\lambda D_\mu U^\dagger + {\lambda^\dagger} D_\mu U}\vev{{\lambda_L} D^\mu U^\dagger U +{\lambda_R} D^\mu U U^\dagger} + \ldots ,
\end{equation}
where each LEFT operator corresponds to a string of chiral operators of increasing dimension in the low-momentum
expansion, and we only display one example term.  The matrices $\lambda^{(\dagger)}$, 
$\lambda_{L,R}$ need to be chosen in accordance with the flavor structure $\psi$, $\chi$ of the four-fermion operator.

\section{\boldmath{Dalitz plot asymmetries in $\eta\to\pi^0\pi^+\pi^-$} and related decays}\label{sec:eta3pi}

Testing $C$-odd interactions via interference effects (instead of via rates of decays that are strictly possible only if $C$ is violated) is particularly favorable for decays with large branching fractions.  For this reason, it has been suggested to search for $C$ violation in Dalitz plot asymmetries in $\eta\to\pi^0\pi^+\pi^-$~\cite{Gardner:2019nid}, one of the dominant $\eta$ decay modes.   Analogous suggestions have subsequently also been made for $\eta'$ three-body decays, $\eta'\to\eta\pi^+\pi^-$ and $\eta'\to\pi^0\pi^+\pi^-$, with the rationale that these are partly sensitive to underlying short-distance operators of different isospin~\cite{Akdag:2021efj}.

The possible symmetry breaking patterns in $\eta\to\pi^0\pi^+\pi^-$ are most easily understood by noting that it manifestly breaks $G$-parity.  As the strong and electromagnetic interactions preserve charge conjugation invariance, isospin breaking is the source thereof in the SM; with electromagnetic effects strongly suppressed~\cite{Sutherland:1966zz,Baur:1995gc,Ditsche:2008cq}, this makes $\eta\to3\pi$ an ideal process to extract information on the light quark mass difference $m_u-m_d$~\cite{Colangelo:2018jxw} (cf.\ also references therein),
far less affected by electromagnetic effects than the extraction from meson masses (see Ref.~\cite{Stamen:2022uqh} for recent work on the latter).
Once we allow for BSM effects, two additional amplitudes that break $C$-invariance can be added, either conserving or breaking isospin, such that the full decay amplitude is written as 
\begin{equation}
\label{eq:FullAmpEta3Pi}
    \M_c(s,t,u)= \M^{\not C}_0(s,t,u) + \xi\,\M^C_1(s,t,u) + \M^{\not C}_2(s,t,u) \,, \qquad \xi = \frac{\big(M_{K^+}^2- M_{K^0}^2\big)_{\text{QCD}}}{3\sqrt{3}F^2_\pi} \,,
\end{equation}
where the subscripts denote the total isospin of the three-pion final state, and $\xi$ parametrizes the isospin breaking in the SM amplitude.  The three amplitudes can individually be analyzed in terms of Khuri--Treiman equations~\cite{Khuri:1960zz}.  These integral equations resum pion--pion rescattering effects in the final state to all orders.  To this end, the amplitudes are decomposed into so-called single-variable amplitudes (SVAs) according to reconstruction theorems~\cite{Kambor:1995yc,Anisovich:1996tx,Gardner:2019nid} (cf.\ also Ref.~\cite{Bernard:2024ioq} for analogous decompositions in $K\to3\pi$ decays) up to (and including) $P$-waves:
\begin{align}
    \M^C_1(s,t,u)&=\F_0(s)+(s-u)\,\F_1(t)+(s-t)\,\F_1(u)+\F_2(t)+\F_2(u)-\frac{2}{3}\F_2(s)\,, \notag    \\
    \M_0^{\not C}(s,t,u)&=(t-u)\,\G_1(s) + (u-s)\,\G_1(t)+(s-t)\,\G_1(u)\,, \notag   \\[1mm]
    \M_2^{\not C}(s,t,u)&=2(u-t)\,\H_1(s) +(u-s)\,\H_1(t)+(s-t)\,\H_1(u) -\H_2(t)+\H_2(u)\,,
	\label{eq:reconstruction-theorems}
\end{align}
where the subscripts denote the corresponding two-pion isospin (which in turn identifies the angular momentum uniquely via Bose symmetry).  While $\M^C_1(s,t,u)$ is even under the exchange $\pi^+ \leftrightarrow \pi^-$ that corresponds to $t\leftrightarrow u$, $\M_{0,2}^{\not C}$ are odd. The SVAs can be written as inhomogeneous Omn\`es solutions
\begin{equation}
    \label{eq:DispersionIntegral}
    \A_I(s)= \Omega_I(s)\,\bigg(P_{n-1}(s)+\frac{s^n}{\pi}\int_{4M_\pi^2}^\infty \frac{\diff x}{x^{n}}\, \frac{\sin\delta_I(x)\,\hat{\A}_I(x)}{|\Omega_I(x)|\,(x-s)}\bigg)\,,
\end{equation}
$\A\in\{\F,\G,\H\}$,
where the Omn\`es functions $\Omega_I(s)$ are given in terms of the corresponding pion--pion phase shifts $\delta_I(s)$, and the subtraction polynomials $P_{n-1}(s)$ comprise the free parameters of the dispersion-theoretical amplitude representation.  It was found~\cite{Akdag:2021efj} that a minimal subtraction scheme for $\M^C_1$ depends on three (real) constants; it allows us to fit data for $\eta\to\pi^+\pi^-\pi^0$~\cite{Anastasi:2016cdz} and $\eta\to3\pi^0$~\cite{Prakhov:2018tou} very well, as well as to fulfill constraints from chiral perturbation theory at $\mathcal{O}\big(p^4\big)$~\cite{Gasser:1984pr,Colangelo:2018jxw}.
Employing strictly analogous assumptions on the amplitude behavior, justified phenomenologically for the SM part, it can be shown~\cite{Akdag:2021efj} that $\M_{0,2}^{\not C}$ both depend on one single (real) subtraction constant each, which can be matched unambiguously onto leading Taylor invariants:
\begin{equation}
    \M^{\not C}_0(s,t,u)=i\,g_0\,(s-t) (u-s) (t-u)+\mathcal{O}\big(p^8\big) \,, \qquad
    \M^{\not C}_2(s,t,u)=i\,g_2\,(t-u)+\mathcal{O}\big(p^4\big) \,. \label{eq:eta3pi-Codd-tree}
\end{equation}
While $\M^{\not C}_2$ starts at $\O(p^2)$ in the chiral expansion, $\M^{\not C}_0$ is, for symmetry reasons, strongly suppressed to $\O(p^6)$~\cite{Akdag:2021efj,Shi:2024yfa}.

It is important to understand precisely how the generation of $C$-odd observables
works as an interference of SM and BSM, or $C$-even and $C$-odd amplitudes.  Obviously,
any such signal is proportional to
\begin{equation}
\sum_{I=0,2} \Re \Big( \M^C_1 \M^{\not C *}_I + \M^{C*}_1 \M^{\not C}_I \Big) \,.  \label{eq:interference}
\end{equation}
We therefore conclude the following:
\begin{enumerate}
\item
As hermiticity of the amplitudes enforces the relative factor of $i$ in the $C$-odd amplitudes,
cf.\ Eq.~\eqref{eq:eta3pi-Codd-tree}, compared to the SM decay amplitude, no Dalitz plot
asymmetry can be generated at tree level, as the interference~\eqref{eq:interference} vanishes.
\item
  As a result, in strict analogy to what is well known as a requirement to cause $CP$ rate
  asymmetries in the weak interactions, only different strong phases generate $C$-odd asymmetries
  in $\eta\to\pi^0\pi^+\pi^-$.  Because the observable effect is proportional to such phase
  differences, it seems advisable to employ the pion--pion phases exactly, rather than approximate
  them perturbatively in one-loop chiral perturbation theory, as this is know to be of insufficient
  accuracy phenomenologically to describe the Dalitz plot in the SM~\cite{Gasser:1984pr}.
\item
  Formally, if we rewrite all amplitudes in the chiral expansion,
  \begin{align}
    \M^C_1 &= \M^{C(2)}_1 + \M^{C(4)}_1 + \O\big(p^6\big) \,, \notag\\
    \M^{\not C}_2 &= i \Big (\tilde\M^{\not C(2)}_2 + \tilde\M^{\not C(4)}_2 \Big) + \O\big(p^6\big) \,, \qquad
    \M^{\not C}_0 = \O\big(p^6\big) \,, 
  \end{align}
  where the additional superscripts in brackets denote the chiral order, and
  both $\M^{C(2)}_1$ and $\tilde\M^{\not C(2)}_2$ are real, the interference term can be approximated as
  \begin{equation}
    \sum_{I=0,2} \Re \Big( \M^C_1 \M^{\not C *}_I + \M^{C*}_1 \M^{\not C}_I \Big)
    = 2 \tilde\M^{\not C(2)}_2 \Im \M^{C(4)}_1 - 2 \M^{C(2)}_1 \Im \tilde\M^{\not C(4)}_2 + \O\big(p^8) \,.
  \end{equation}
  We see that the leading interference involves both the imaginary part of the well-known SM one-loop amplitude~\cite{Gasser:1984pr} and the one of the one-loop $C$-odd $I=2$ amplitude.  Omitting one of the two
  terms, as suggested in Ref.~\cite{Shi:2024yfa}, is not justified by the chiral expansion.
\end{enumerate}

The effective coupling constants $g_{0,2}$ can be constrained from KLOE Dalitz plot data~\cite{Anastasi:2016cdz},
$g_0 = -2.8(4.5) \GeV^{-6}$, $g_2 = -9.3(4.6) \times 10^{-3} \GeV^{-2}$,
which restricts $C$ violation to the permille level.  While the kinematical dependence of the two $C$-odd amplitudes and their different interference patterns can clearly not be resolved (all $C$-/$CP$-violating signals vanish within $1$--$2\sigma$), the small phase space severely reduces the sensitivity to the isoscalar amplitude $\M^{\not C}_0$~\cite{Gardner:2019nid,Akdag:2021efj}: the natural theoretical expectation from chiral power counting would be $|g_0/g_2| \sim 1 \GeV^{-4}$.

In principle, this kinematic suppression could be partially overcome in the decay $\eta'\to\pi^0\pi^+\pi^-$, in which one would expect $C$ violation to be caused by the same fundamental operators; indeed, the relative theoretical sensitivity to $g_0$ would be enhanced by about two orders of magnitude compared to $\eta\to\pi^0\pi^+\pi^-$~\cite{Akdag:2021efj}.  However, as this is a comparably rare decay ($\mathcal{B}(\eta'\to\pi^0\pi^+\pi^-) \approx 3.6\times10^{-3}$), an analysis of $C$ violation based on the existing data by BESIII~\cite{BESIII:2016tdb} is currently not too promising.

This is different for $\eta'\to\eta\pi^+\pi^-$~\cite{BESIII:2017djm}, for which a Khuri--Treiman analysis of the SM amplitude is available already~\cite{Isken:2017dkw}.  In this case, the SM decay conserves isospin, while a $C$-odd contribution would change the isospin by 1.  Charge asymmetries in the $\eta'\to\eta\pi^+\pi^-$ Dalitz plot therefore test a BSM operator of different isospin, independent of the ones constrained in $\eta\to\pi^0\pi^+\pi^-$.  A dispersive analysis of the corresponding amplitudes has also been performed~\cite{Akdag:2021efj}, and while the resulting constraints are not quite as strong yet, $C$ violation is also restricted to the percent level in this channel. 

\section{\boldmath{Correlating $\eta\to\pi^0\pi^+\pi^-$ and $\eta\to\pi^0\ell^+\ell^-$}}\label{sec:eta-pi0ll}

\begin{figure}
    \centering
    \includegraphics[width=0.9\linewidth]{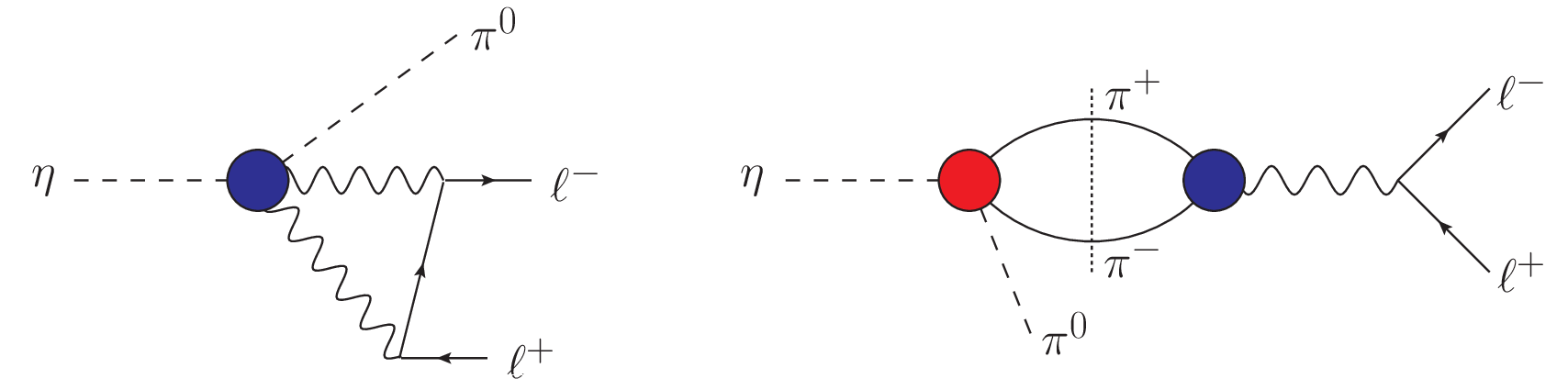}
    \caption{$C$-conserving two-photon mechanism for the decay $\eta\to\pi^0\ell^+\ell^-$ (left), and discontinuity of the $C$-odd $\eta\to \pi^0 \gamma^*$ transition form factor (right), given by the $C$-odd $\eta\pi^0\to\pi^+\pi^-$ $P$-wave (red) and the pion vector form factor (blue).}
    \label{fig:eta-pi0ll}
\end{figure}

Of the $C$-odd $\eta$ decay modes listed in Table~\ref{tab:CVdecays}, the seemingly simplest, $\eta\to\pi^0\gamma^{(*)}$,
is complicated by the fact that, as a consequence of gauge invariance and angular momentum conservation alone, it is still forbidden for real photons.  
This can be seen from the corresponding matrix element
\begin{equation}
\langle \pi^0(k) | j_\mu(0) | \eta(p) \rangle = -i \big[ q^2(p+k)_\mu - (M_\eta^2-M_\pi^2) q_\mu \big] F_{\eta\pi^0}(q^2) \,
\end{equation}
(cf.\ the similar, although flavor-changing, decays $K\to\pi\ell^+\ell^-$~\cite{DAmbrosio:1998gur,Kubis:2010mp}), which vanish for real photons $q^2=0$.  Potential nonvanishing, $C$-odd transitions can therefore only be assessed in the corresponding dilepton decays $\eta\to\pi^0\ell^+\ell^-$, for which, however, a $C$-conserving two-photon exchange forms an irreducible SM background; cf.\ Fig.~\ref{fig:eta-pi0ll} (left).
The corresponding SM branching ratios have been recalculated recently~\cite{Schafer:2023qtl},
$\mathcal{B}(\eta\to\pi^0e^+e^-) = 1.36(15)\times 10^{-9}$ and
$\mathcal{B}(\eta\to\pi^0\mu^+\mu^-) = 0.67(7)\times 10^{-9}$ 
(as well as similar orders of magnitude for $\eta'\to\pi^0\ell^+\ell^-$ and $\eta'\to\eta\ell^+\ell^-$), 
based on a vector-meson-dominance model for the corresponding two-photon decays and phenomenologically viable transition
form factors.  
These are to be compared to the current experimental upper limits, $\mathcal{B}(\eta\to\pi^0e^+e^-) < 7.5\times 10^{-6}$~\cite{WASA-at-COSY:2018jdv} and
$\mathcal{B}(\eta\to\pi^0\mu^+\mu^-) < 5\times 10^{-6}$~\cite{Dzhelyadin:1980ti}.
Evidence for a $C$-odd signal can hence only be claimed if measured branching ratios significantly exceed the theoretical SM prediction, or in the nowadays rather unlikely case that interference effects between $C$-even and $C$-odd mechanisms allow us to observe Dalitz plot asymmetries in differential decay distributions.

Long-distance contributions to the isovector component of the transition form factor $F^{(1)}_{\eta\pi^0}$ can be reconstructed
dispersively, based on the two-pion intermediate state [cf.\ Fig.~\ref{fig:eta-pi0ll} (right)], according to~\cite{Akdag:2023pwx}
\begin{equation}
F^{(1)}_{\eta\pi^0}(q^2) = \frac{i}{48\pi^2}\int_{4M_\pi^2}^{\infty}\diff x\, \sigma^3_\pi(x) F_\pi^{V*}(x) \,\frac{f_{\eta\pi^0}(x)}{x-q^2} \,,
\end{equation}
where $F_\pi^V$ is the pion vector form factor, and $f_{\eta\pi^0}$ denotes the $C$-odd $P$-wave projection of the
$\eta\pi^0\to\pi^+\pi^-$ amplitude.  This relation is analogous to the one between the $\omega\to\pi^0$ transition
form factor and the $\omega\to3\pi$ decay amplitude in the SM~\cite{Schneider:2012ez}.  Continuation to the second Riemann
sheet of the complex $q^2$ plane allows us to extract $C$-odd $\rho$-meson coupling constants and relate them to the
$\eta\to\pi^0\pi^+\pi^-$ coupling constants $g_{0,2}$ in a model-independent way.  Isoscalar long-distance contributions
$F^{(0)}_{\eta\pi^0}$ due to $\omega$-exchange can then be estimated using $SU(3)$ symmetry and vector-meson dominance.  
\begin{table}
\begin{center} \renewcommand{\arraystretch}{1.3}
  \begin{tabular}{llll}
    \toprule
    & isovector & isovector + isoscalar & experiment \\ \midrule
$\BR(\eta\to\pi^0e^+e^-)$ & $< 20 \cdot 10^{-6}$ & \hspace*{4mm} $ < 29 \cdot 10^{-6}$ & $< 7.5 \cdot 10^{-6}$~\cite{WASA-at-COSY:2018jdv} \\
    $\BR(\eta\to\pi^0\mu^+\mu^-)$ & $< 7.2 \cdot 10^{-6}$ & \hspace*{4mm} $ < 10 \cdot 10^{-6}$ & $< 5.0 \cdot 10^{-6}$~\cite{Dzhelyadin:1980ti} \\
    \bottomrule
    \end{tabular}
\end{center}
\caption{Predicted upper limits on$\eta\to\pi^0\ell^+\ell^-$ branching ratios, due to
  isovector as well as isovector plus isoscalar $C$-odd long-distance  contributions, compared to the experimental upper limits.
\label{tab:BRs-eta-pill}}
\end{table}
The resulting theoretical limits on the $\eta\to\pi^0\ell^+\ell^-$ branching ratios, based on the $\eta\to3\pi$ Dalitz plot
analysis, are summarized in Table~\ref{tab:BRs-eta-pill}.  We note that the experimental upper limits are already more
rigorous than the theoretical estimates; this is largely due to the fact that the isoscalar coupling $g_0$ is only badly
constrained from the Dalitz plot analysis, see the discussion in the previous section, but is not similarly suppressed
in its impact on the transition form factor.  As a result, this relation can be used to constrain $|g_0|$ (slightly) more rigorously.  Similar relations have also been worked out for $C$-odd $\eta'$ transition form factors~\cite{Akdag:2023pwx}, where, however,
interferences with short-ranged effects such as direct $C$-odd photon couplings or leptonic operators have been neglected
so far.

\section{\boldmath{Dalitz plot asymmetries in $\eta^{(\prime)}\to\gamma\ell^+\ell^-$}}\label{sec:eta-gll}

Within the Standard Model, the decays $\eta^{(\prime)}\to\gamma\ell^+\ell^-$ are described in terms of the
(singly-virtual) transition form factors $F_{\eta^{(\prime)}\gamma^*\gamma}(q^2) \equiv F_{\eta^{(\prime)}\gamma^*\gamma^*}(q^2,0)$, which are in general defined via the matrix element
\begin{equation}
i \int \diff^4x \, e^{iq_1x} \, i \left\langle 0 \left| T \big\{ j_\mu(x)j_\nu(0) \big\} \right| \eta^{(\prime)} (q_1+q_2)\right\rangle
= \epsilon_{\mu\nu\alpha\beta} q_1^\alpha q_2^\beta \, F_{\eta^{(\prime)}\gamma^*\gamma^*}(q_1^2,q_2^2) \,. \label{eq:defTFF}
\end{equation}
These have been extensively studied dispersively in the context of hadronic light-by-light contributions
to the muon's anomalous magnetic moment
(see, e.g., Refs.~\cite{Holz:2024lom,Holz:2024diw} and references therein).
Similar to the case of $\eta\to\pi^0\pi^+\pi^-$, $C$-odd effects could be observed by studying charge asymmetries in the Dalitz plot distribution of the three-particle final state; again, this observable is an interference
effect between (SM) $C$-even and (BSM) $C$-odd amplitudes.

A $C$-odd contribution to this decay 
can be induced by the dimension-7 quark--lepton operator
\begin{equation}
\O_{\ell\psi}^{(d)} \equiv  c_{\ell\psi}^{(d)}\, \bar\ell \gamma^\mu \gamma_5 \ell \, \bar\psi \dvec {D}_\mu \gamma_5 \psi \,,
\end{equation}
where also alternative terms appearing at dimension~8 have been investigated~\cite{Herz2023}.  The leading chiral
realization of $\O_{\ell\psi}^{(d)}$ occurs at $\O(p^4)$, a possible operator being given by
\begin{equation}
        X_{\ell\psi}^{(d)}= \frac{v}{\Lambda^4}g_1^{(d)}c^{(d)}_{\ell\psi} \, i \, \vev{\left(\lambda D_\nu \bar U^\dagger f^\munu_R-\lambda^\dagger D_\nu \bar U f^\munu_L\right)-\left(f^\munu_R D_\nu \bar U \lambda^\dagger-f^\munu_L D_\nu \bar U^\dagger \lambda\right)} \, \bar\ell \gamma_\mu\gamma_5 \ell \,.
\end{equation}
This leads to a $C$-odd BSM matrix element for $\eta^{(\prime)} \to \gamma(\lambda,q) \ell^+(p_+)\ell^-(p_-)$ of the form
\begin{equation}
  \mathcal{M}_\text{BSM} \propto (p_-+p_+)_\nu \left[\bar{u}(p_-)\gamma_\mu\gamma^5v(p_+) \right]  \left(q^\mu\varepsilon(\lambda)^{\dagger\nu}-q^\nu\varepsilon(\lambda)^{\dagger\mu} \right)\, .
\end{equation}

\begin{figure}
  \centering
  \includegraphics*[width=\linewidth]{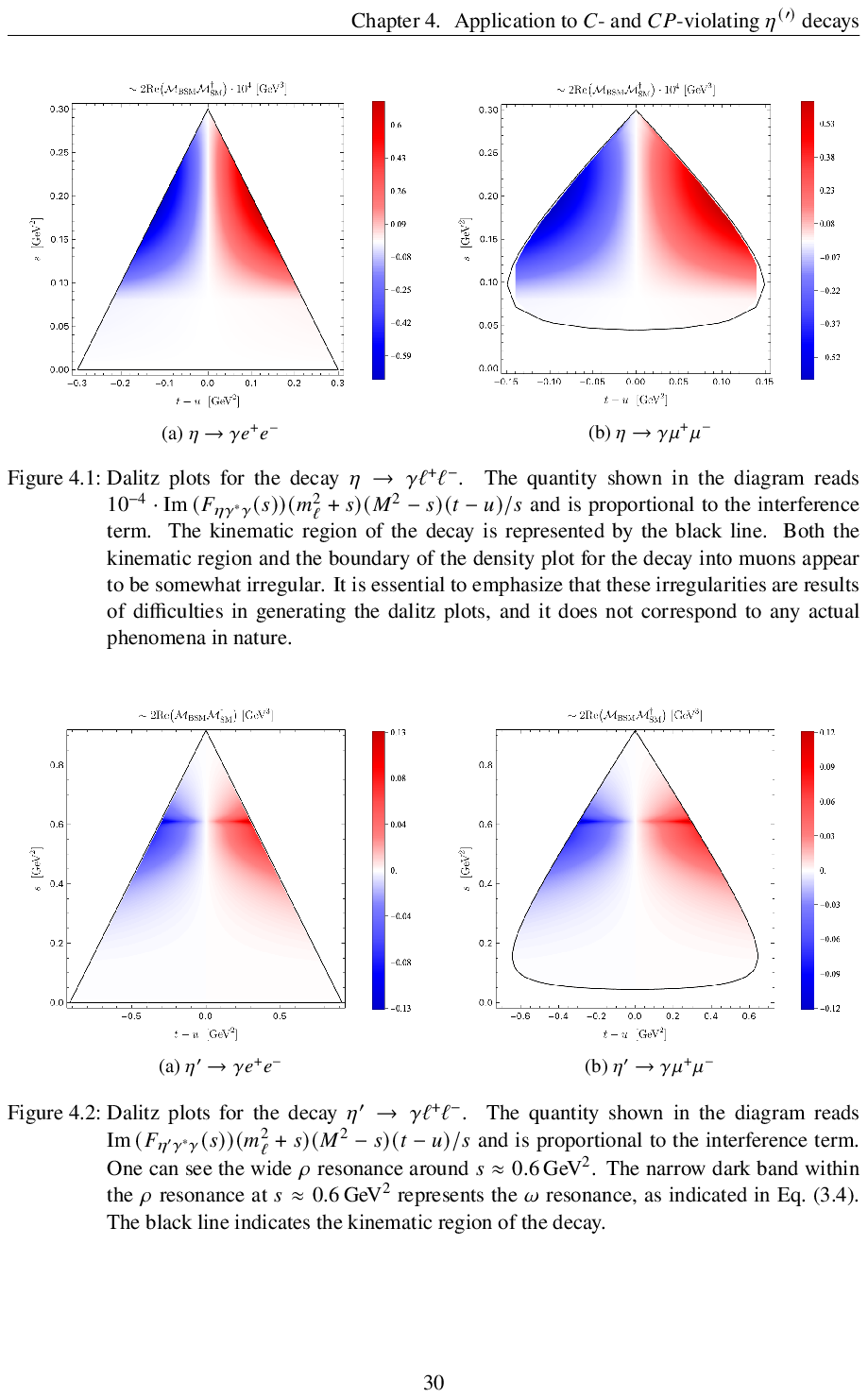}\\
  \includegraphics*[width=\linewidth]{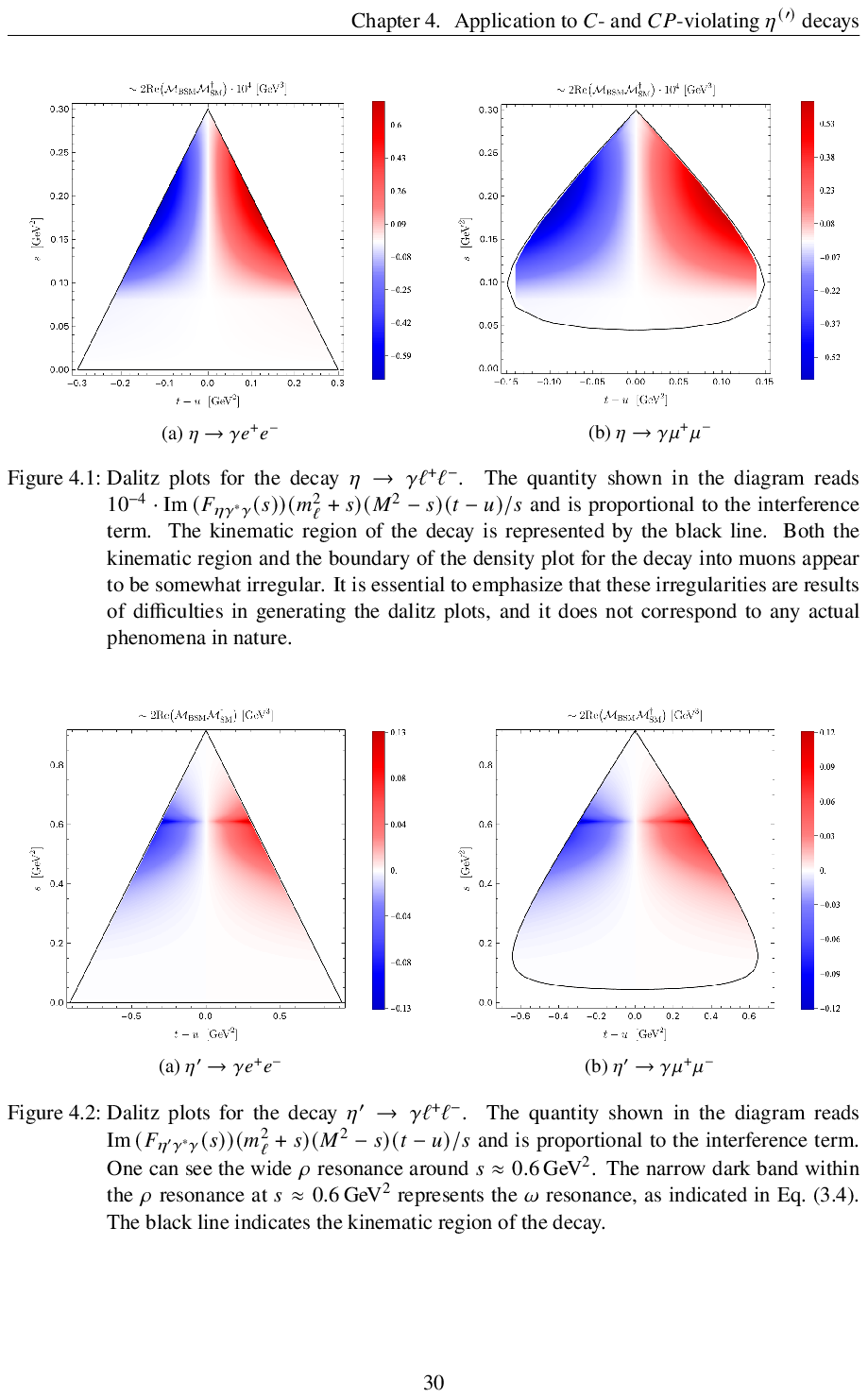}
  \caption{Dalitz plot asymmetries for $\eta\to\gamma e^+e^-$ (top left), $\eta\to\gamma \mu^+\mu^-$ (top right), $\eta'\to\gamma e^+e^-$ (bottom left), and $\eta'\to\gamma \mu^+\mu^-$ (bottom right). Figures taken from Ref.~\cite{Herz2023}.  \label{fig:asym-eta-gll}
  }
\end{figure}

Observable Dalitz plot asymmetries are again generated by the interference of $C$-odd with $C$-even amplitudes,
which, once more, would vanish for tree-level amplitudes with no additional final-state-interaction phase.
As in this case, the $C$-odd operator is local, such interferences scale with the imaginary parts of the
transition form factors~\cite{Herz2023},
\begin{equation}
  \Re\big(\M_\text{BSM} \M_\text{SM}^\dagger \big) \propto \Im F_{\eta^{(\prime)}\gamma^*\gamma}(q^2) \,.
\end{equation}
Because these are precisely the basis of dispersion-theoretical analyses, they are very well
understood~\cite{Hanhart:2013vba,Kubis:2015sga,Holz:2022hwz}; what is more, in this context only the
imaginary parts accessible within the physical decay region are relevant and not those of higher intermediate
states necessary for a complete reconstruction of the real part~\cite{Holz:2024lom,Holz:2024diw}.
As a result, the imaginary parts are dominated by the two-pion intermediate state; the three-pion spectral
function can be approximated to excellent precision by the narrow $\omega(782)$ resonance, which only plays
a role for the $\eta'$ decay:
\begin{equation}
  \Im F_{\eta^{(\prime)}\gamma^*\gamma}(q^2)  = \frac{q^2}{96\pi} \left(1-\frac{4M_\pi^2}{q^2}\right)^{3/2}
  \big(F_\pi^V(q^2)\big)^* f_1^{\eta^{(\prime)}}(q^2)
  + \frac{F_{\etap\gamma\gamma}w_{\etap\omega\gamma} M_\omega^3 \Gamma_\omega}{(M_\omega^2 - q^2)^2 + M^2_\omega\Gamma^2_\omega}
  \,,
\end{equation}
where $f_1^{\etap}(q^2)$ denotes the $P$-wave projection of
the $\etap \to \pi^+\pi^-\gamma$ amplitude, $F_{\etap\gamma\gamma} = F_{\etap\gamma^*\gamma^*}(0,0)$,
and $w_{\etap\omega\gamma}$ are weight factors for the $\omega$ pole that can be extracted from
$\omega\to\eta\gamma$ and $\eta'\to\omega\gamma$~\cite{Gan:2020aco}.
The resulting contributions to the four possible Dalitz plot asymmetries are depicted in Fig.~\ref{fig:asym-eta-gll}, where the different scaling of the $\eta$ and the $\eta'$ decays needs to be noted.  We observe the following
consequences of these asymmetries being proportional to the transition form factors' imaginary parts~\cite{Herz2023}:
\begin{enumerate} 
\item As both the $\rho(770)$ and the $\omega(782)$ resonances occur within the physical decay region
  of the $\eta'$ decays, while only the (nonresonant) low-energy tail of the pion--pion continuum provides
  an imaginary part in the case of the $\eta$, asymmetries for the $\eta' \to \gamma\ell^+\ell^-$ decays
  are larger by several orders of magnitude.
\item
  As the SM amplitude is dominated by the photon pole, and hence the Dalitz decay with an $e^+e^-$ final state
  by very low dilepton invariant masses (the region below the $\pi^+\pi^-$ threshold where the transition form
  factor is real), the asymmetries are also significantly larger for the dimuon final state than for the
  electron--positron one.
\item
  Finally, note that no such $C$-odd  asymmetries can be generated for the Dalitz decay of the neutral pion $\pi^0 \to \gamma e^+e^-$,
    as in that case, the dilepton masses always stays below the $\pi^+\pi^-$ cut.
\end{enumerate}
There is, therefore, a clear hierarchy in sensitivity between the different $P\to\gamma\ell^+\ell^-$ decays
when searching for $C$-odd effects in the Dalitz plot distributions.

\section{\boldmath{Asymmetries in $\etap\to\pi^+\pi^-\ell^+\ell^-$}}\label{sec:eta-pipill}
\begin{figure}
  \centering
    \includegraphics*[width=0.55\linewidth]{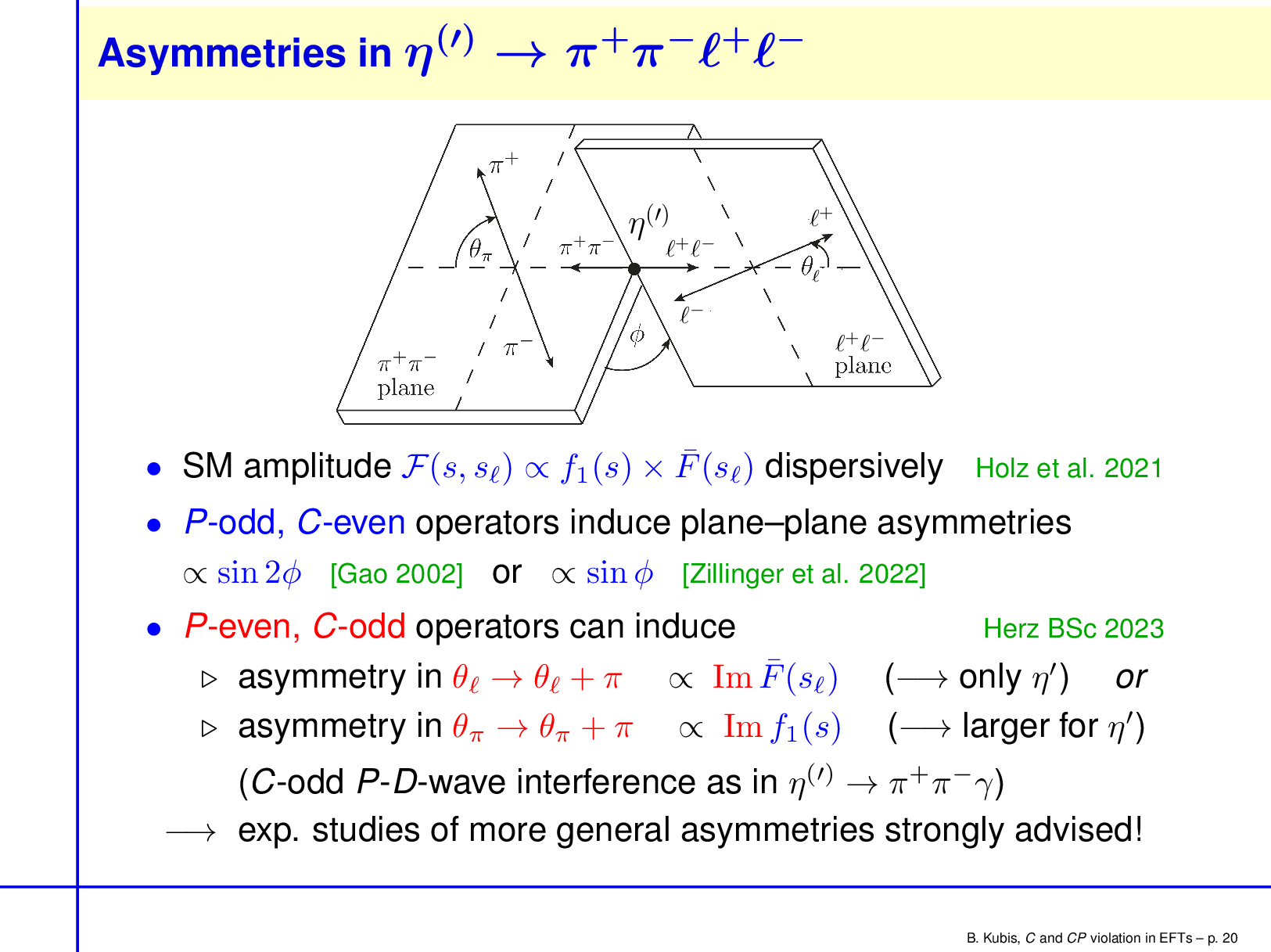}
    \caption{Kinematical variables for the semileptonic four-body decays $\etap\to\pi^+\pi^-\ell^+\ell^-$.
    \label{fig:eta-kinematics}}
\end{figure}

The richer decay kinematics of the semileptonic four-body decays $\etap\to\pi^+\pi^-\ell^+\ell^-$, cf.\
Fig.~\ref{fig:eta-kinematics}, allow for a multitude of different discrete symmetry tests.
It has been suggested a long time ago to search for $CP$~violation via
asymmetries in the angular distribution $\phi$ between the dipion and the dilepton planes,
based on a $P$-odd, $C$-even four-quark operator~\cite{Gao:2002gq} that induces a $\sin2\phi$ dependence.
This can, however, be very rigorously constrained indirectly by EDM
measurements~\cite{Gan:2020aco}, way beyond what will be testable experimentally in the foreseeable
future~\cite{KLOE:2008arm}.  Alternative, scalar, $P$-odd, $C$-even operators that evade EDM constraints
to some extent~\cite{Sanchez-Puertas:2018tnp} induce a different $CP$-odd signal
$\propto \sin\phi$~\cite{Zillinger:2022eva}, which also ought to be tested and constrained in future
high-statistics $\eta$ and $\eta'$ decay facilities.  Asymmetries alternatively proportional to
$\sin2\phi$ and $\sin\phi$ are known to correspond to indirect and direct $CP$~violation in
related $K_L$ decays~\cite{Heiliger:1993qt}.

The alternative symmetry breaking pattern investigated in this article, based on $P$-even, $C$-odd
interactions, add yet more possibilities to define and investigate $\etap\to\pi^+\pi^-\ell^+\ell^-$
decay asymmetries~\cite{Herz2023}.
To begin with, the SM decay amplitude, traditionally parametrized in terms of vector-meson-dominance
models, is now also available in a dispersive representation~\cite{Zillinger:2022eva}.
At low energies relevant for decay kinematics, the dependence on the dipion ($s$) and dilepton ($s_\ell$)
invariant masses squared factorizes to very good precision~\cite{Holz:2015tcg},
\begin{equation}
\F(s,s_\ell) \propto f_1^{\etap}(s) \bar F(s_\ell) \,,
\end{equation}
where $f_1^{\etap}(s)$ denotes the same $\etap\to\pi^+\pi^-\gamma$ $P$-wave amplitude as in the previous section,
and $\bar F(s_\ell)$ was parametrized as a dipole form based on $\rho$ and $\rho'$ resonance contributions
in Ref.~\cite{Zillinger:2022eva}; cf.\ Refs.~\cite{Holz:2024lom,Holz:2024diw} for more elaborate schemes.
Both $f_1^{\etap}(s)$ and $\bar F(s_\ell)$ have branch cuts in their respective variables starting at $4M_\pi^2$.

$C$-odd quark--lepton operators now can induce different types of asymmetries~\cite{Herz2023}:
\begin{enumerate}
\item A $C$-odd lepton asymmetry $\theta_\ell \leftrightarrow \theta_\ell + \pi$ requires interference with a
  nonvanishing $\Im \bar F(s_\ell)$, which is only allowed for the $\eta'$ decays, as only here, the
  two-pion threshold in the dilepton invariant mass can be reached, which is equivalent
  to the four-(charged-)pion decays being kinematically allowed~\cite{Guo:2011ir}.
\item Alternatively, the $C$-odd pion asymmetry $\theta_\pi \leftrightarrow \theta_\pi + \pi$ is generated
  by $\Im f_1^{\etap}(s)$, which, for the same reason discussed in Sec.~\ref{sec:eta-gll}, is much larger
  for $\eta'$ decays, hence resulting in a much increased sensitivity to the underlying BSM operators.
  In fact, this last mechanism is the same as the $C$-odd $P$--$D$-wave interference that can also
  be tested with a real photon in the final state in $\etap \to \pi^+\pi^-\gamma$~\cite{Barrett:1966,Akdag:2022sbn}.
\end{enumerate}
Our conclusion is that future experiments ought to be open-minded to test as many different symmetry-breaking
BSM scenarios in parallel as possible, without making model assumptions that dismiss some of the scenarios
discussed here out of hand.

\section{Summary}

Decays of $\eta$ and $\eta'$ mesons allow for a vast range of different physics investigations, both within the Standard Model and beyond.
Here we have focused on possible new patterns of discrete symmetry violation, in particular
those that violate $CP$ by breaking $C$, but preserving $P$,  
as electric dipole moments set very rigorous limits on $P$ and $CP$ violation in $\ep$ decays.
As a recurring theme, we have emphasized the importance of strong final-state-interaction phases:
because BSM effects are expected to be small by all means, it is advantageous to search for SM--BSM
interference effects through asymmetry measurements, rather than for rates of $C$-odd processes alone
that will be doubly suppressed in small BSM parameters.
SM--BSM interferences, however, necessarily require imaginary parts due to rescattering to be observable.

We have established a dispersive framework to investigate charge asymmetries in
$\ep\to\pi^0\pi^+\pi^-$ and $\eta'\to\eta\pi^+\pi^-$ Dalitz plots, which take all pairwise rescattering
of final-state mesons into account via the Khuri--Treiman formalism.  Subtraction constants
can be matched onto chiral perturbation theory in a consistent manner, and it is understood how to select
the relevant number of independent parameters also for the BSM amplitudes.  It is known that a perturbative
treatment of final-state interactions is insufficient phenomenologically for $\eta\to3\pi$, and it is
necessary to treat such rescattering effects consistently in the interfering SM and BSM amplitudes.

Dispersion relations allow us to relate the $C$-odd $\eta\to\pi^0\pi^+\pi^-$ amplitude to long-distance
effects in the single-photon $\eta\to\pi^0\ell^+\ell^-$ decays, fixing the isovector part unambiguously.
Flavor-symmetry arguments and vector-meson dominance
help to also constrain the isoscalar contributions to some extent.
These decays also have a SM, $C$-even, background
from two-photon exchange, which is, however, loop suppressed and very small.
In some cases, constraints from these semileptonic decays are competitive with those extracted from
the hadronic Dalitz plot asymmetries.

Other semileptonic $\ep$ decays have also been investigated for potential $C$-odd observables, in particular
$\ep\to\gamma\ell^+\ell^-$ and $\ep\to\pi^+\pi^-\ell^+\ell^-$.  The SM mechanisms for these decays are
well-studied due to their relevance to hadronic contributions
to $(g-2)_\mu$; in particular the imaginary parts of the corresponding transition form factors,
which once more are prerequisites for
nonvanishing $C$-odd interference effects, are extremely tightly fixed in a model-independent
way.  The theoretical knowledge on these SM form factors ought to be used as a leverage to constrain
BSM operators with the best possible accuracy.  For reasons of phase space and resonant enhancement
of the imaginary parts involved, $\eta'$ decays show much larger sensitivities in this respect than their $\eta$ counterparts.

New experimental results from facilities such as the JLab Eta Factory~\cite{Gan:2020aco}, REDTOP~\cite{REDTOP:2022slw}, or a super $\eta$ factory at HIAF~\cite{Chen:2024wad}, making optimal use of these theoretical advances, are eagerly awaited.

\section*{Acknowledgements}
I am indebted to my collaborators 
H.~Akdag,
J.~Herz,
T.~Isken,
P.~S\'anchez-Puertas,
A.~Wirzba,
and M.~Zillinger
for collaborations on the different projects summarized in these proceedings.

\bibliographystyle{JHEP_mod}
\bibliography{Literature,base}

\end{document}